\documentclass[10pt]{article}

\usepackage[T1]{fontenc}
\usepackage[utf8]{inputenc}
\usepackage[english]{babel}
\usepackage{amsmath,latexsym,amsfonts,amsthm,bm,mathtools}
\usepackage{graphicx}
\usepackage[top=1.5in, bottom=1.5in, left=1.25in, right=1.25in]{geometry}
\usepackage{booktabs,url,tikz}

\title{A Maximum Entropy Procedure to Solve \\Likelihood Equations}
\author{Antonio Calcagn\`{i}\footnote{Corresponding author: antonio.calcagni@unipd.it}, 
	Livio Finos,
	Gianmarco Alto\'{e}, \\and 
	Massimiliano Pastore\\ \\
 University of Padova 
}

\date{}

\DeclareMathAlphabet{\mathcall}{OMS}{zplm}{m}{n}

\begin{document}

\maketitle

\begin{abstract}
In this article we provide initial findings regarding the problem of solving likelihood equations by means of a maximum entropy approach. Unlike standard procedures that require equating at zero the score function of the maximum-likelihood problem, we propose an alternative strategy where the score is instead used as external informative constraint to the maximization of the convex Shannon's entropy function. The problem involves the re-parameterization of the score parameters as expected values of discrete probability distributions where probabilities need to be estimated. This leads to a simpler situation where parameters are searched in smaller (hyper) simplex space. We assessed our proposal by means of empirical case studies and a simulation study, this latter involving the most critical case of logistic regression under data separation. The results suggested that the maximum entropy re-formulation of the score problem solves the likelihood equation problem. Similarly, when maximum-likelihood estimation is difficult, as for the case of logistic regression under separation, the maximum entropy proposal achieved results (numerically) comparable to those obtained by the Firth's Bias-corrected approach. Overall, these first findings reveal that a maximum entropy solution can be considered as an alternative technique to solve the likelihood equation.\\\vspace{0.15cm}

\end{abstract}

\section{Introduction}

Maximum likelihood is one of the most used tools of modern statistics. As a result of its attractive properties, it is useful and suited for a wide class of statistical problems, including modeling, testing, and parameters estimation \cite{cox2006principles,stigler2007epic}. In the case of regular and correctly-specified models, maximum likelihood provides a simple and elegant means of choosing the best asymptotically normal estimators. Generally, the maximum likelihood workflow proceeds by first defining the statistical model which is thought to generate the sample data and the associated likelihood function. Then, the likelihood is differentiated around the parameters of interest by getting the likelihood equations (score), which are solved at zero to find the final estimates. In most simple cases, the maximum likelihood solutions are expressed in closed-form. However, analytic expressions are not always available for most complex problems and researchers need to solve likelihood equations numerically. A broad class of these procedures include Newton-like algorithms, such as the Newton--Raphson, Fisher-scoring, and quasi Newton--Raphson algorithms \cite{tanner2012tools}. However, when the sample size is small, or when the optimization is no longer convex as in the case of more sophisticated statistical models, the standard version of Newton--Raphson may not be optimal. In this case, robust versions should instead be used \cite{commenges2006newton}. A typical example of such a situation is the logistic regression for binary data, where maximum likelihood estimates may no longer be available, for instance, when the binary outcome variable can be perfectly or partially separated by a linear combination of the covariates \cite{albert1984existence}. As a result, the Newton--Raphson is unstable with inconsistent or infinite estimates. Other examples include small sample sizes, large numbers of covariates, and multicollinearity among the regressor variables \cite{shen2008solution}. {Different proposals have been made to solve these drawbacks, many of which are based on iterative adjustments of the Newton--Raphson algorithm (e.g., see \cite{firth1993bias,kenne2017median}), penalized maximum likelihood (e.g., see \cite{gao2007asymptotic}), or the homotopy-based method (e.g., see \cite{abbasbandy2007newton}). Among them, bias-corrected methods guarantee the existence of finite maximum likelihood estimates by removing first-order bias \cite{cordeiro1991bias}, whereas homotopy Newton-Raphson algorithms, which are mostly based on Adomian's decomposition, ensure more robust numerical convergences in finding roots of the score function (e.g., see \cite{wu2005study}).}

{Maximum entropy (ME)-based methods have a long history in statistical modeling and inference (e.g., for a recent review see \cite{golan2017foundations}). Since the seminal work by \cite{golan1994recovering}, there have been many applications of maximum entropy to the problem of parameter estimation in statistics, including autoregressive models \cite{golan1996maximumA}, multinomial models \cite{golan1996maximumB}, spatial autoregressive models \cite{marsh2004generalized}, structural equation models \cite{ciavolino2009comparing}, the co-clustering problem \cite{banerjee2007generalized}, and fuzzy linear regressions \cite{ciavolino2016generalized}. What all these works share in common is an elegant estimation method that avoids strong parametric assumptions on the model being used (e.g., error distribution). Differently, maximum entropy has also been widely adopted in many optimization problems, including queueing systems, transportation, portfolio optimization, image reconstruction, and spectral analysis (for a comprehensive review see \cite{kapur1989maximum,fang2012entropy}). In all these cases, maximum entropy is instead used as a pure mathematical solver engine for complex or ill-posed problems, such as those encountered when dealing with differential equations \cite{el2003maximum}, oversampled data \cite{bryan1990maximum}, and data decomposition \cite{calcagni2017analyzing}.}  

The aim of this article is to introduce a maximum entropy-based technique to solve likelihood equations as they appear in many standard statistical models. {The idea relies upon the use of Jaynes' classical ME principle as a mathematical optimization tool \cite{fang2012entropy,el2003maximum,sukumar2004construction}.} In particular, instead of maximizing the likelihood function and solving the corresponding score, we propose a solution where the score is used as the data constraint to the estimation problem. The solution involves two steps: (i) reparametrizing the parameters as discrete probability distributions and (ii) maximizing the Shannon's entropy function w.r.t. to the unknown probability mass points constrained by the score equation. Thus, parameter estimation is reformulated as recovering probabilities in a (hyper) symplex space, with the searching surface being always regular and convex. In this context, the score equation represents all the available information about the statistical problem and is used to identify a feasible region for estimating the model parameters. In this sense, our proposal differs from other ME-based procedures for statistical estimation (e.g., see \cite{golan1996maximum}). {Instead, our intent is to offer an alternative technique to solve score functions of parametric, regular, and correctly specified statistical models, where inference is still based on maximum likelihood theory.}

The reminder of this article is organized as follows. Section \ref{ME_proposal} presents our proposal and describes its main characteristics by means of simple numerical examples. Section \ref{simulation_study} describes the results of a simulation study where the ME method is assessed in the typical case of logistic regression under separation. Finally, Section \ref{discussion} provides a general discussion of findings, comments, and suggestions for further investigations. Complementary materials like datasets and scripts used throughout the article are available to download at \url{https://github.com/antcalcagni/ME-score}, whereas the list of symbols and abbreviations adopted hereafter is available in Table \ref{tab0}.

\begin{table}[!h]
	\caption{List of symbols and abbreviations used throughout the manuscript.}
	\label{tab0}
	\centering
	\begin{tabular}{ll}
		\toprule
		\midrule
		ME & Maximum Entropy\\
		NR & Newton--Raphson algorithm\\
		NFR & Bias corrected Newton--Raphson algorithm\\	
		y & sample of observations\\
		$\mathcall{Y}$ & sample space\\
		$\boldsymbol\theta$ & $J\times 1$ vector of parameters\\
		$\boldsymbol{\hat\theta}$ & estimated vector of parameters\\
		$\boldsymbol{\tilde\theta}$ & reparameterized vector of parameters under ME \\		
		$f(y;\boldsymbol\theta)$ & density function\\
		$l(\boldsymbol\theta)$ & likelihood function\\
		$\mathcall U(\boldsymbol\theta)$, $\mathcall U(\boldsymbol{\tilde\theta})$ & score function\\
		$\mathbf z$ & $K\times 1$ vector of finite elements for $\boldsymbol{\tilde\theta}$\\
		$\mathbf p$ & $K\times 1$ vector of unknown probabilities for $\boldsymbol{\tilde\theta}$\\
		$\mathbf{\hat p}$ & vector of estimated probabilities for $\boldsymbol{\tilde\theta}$\\
		\bottomrule
	\end{tabular}
\end{table}

\section{A maximum entropy solution to score equations}\label{ME_proposal}

Let $\mathbf y=\{y_1,\ldots,y_n\}$ be a random sample of independent observations from the parametric model $\mathcall{M} = \{f(y; \boldsymbol{\theta}): \boldsymbol{\theta} \in \boldsymbol{\Theta}, y \in \mathcall Y\}$, with $f(y; \boldsymbol{\theta})$ being a density function parameterized over $\boldsymbol{\theta}$,  $\boldsymbol{\Theta} \subseteq \mathbb{R}^J$ the parameter space with $J$ being the number of parameters, and $\mathcall Y$ the sample space. Let $${l}(\boldsymbol{\theta}) = \sum_{i=1}^n \ln f(y_i;\boldsymbol \theta)$$ be the log-likelihood of the model and 
$$\mathcall U(\boldsymbol{\theta}) = \nabla_{\boldsymbol{\theta}} l(\boldsymbol{\theta}) = (\partial l/\partial\theta_1,\ldots,\partial l/\partial\theta_j,\ldots,\partial l/\partial\theta_J)$$ the score equation. In the regular case, the maximum likelihood estimate (MLE) $\boldsymbol{\hat\theta}$ of the unknown vector of parameters $\boldsymbol\theta$ is the solution of the score $\mathcall U(\boldsymbol{\theta}) = \mathbf 0_J$. In simple cases, $\boldsymbol{\hat\theta}$ has closed-form expression but, more often, a numerical solution is required for $\boldsymbol{\hat\theta}$, for instance by using iterative algorithms like Newton--Raphson and Expectation-Maximization. 

In the maximum likelihood setting, our proposal is instead to solve $\mathcall U(\boldsymbol{\theta}) = \mathbf 0_J$ by means of a maximum entropy approach (for a brief introduction, see \cite{golan1997maximum}). This involves a two step formulation of the problem, where $\boldsymbol\theta$ is first reparameterized as a convex combination of a numerical support with some predefined points and probabilities. Next, a non-linear programming (NLP) problem is set with the objective of maximizing the entropy of the unknown probabilities subject to some feasible constraints. More formally, let 
\begin{equation}\label{eq1}
\boldsymbol{\tilde\theta} = (\mathbf z_1^T \mathbf p_1,\ldots,\mathbf z_j^T \mathbf p_j,\ldots,\mathbf z_J^T \mathbf p_J)^T
\end{equation}
be the reparameterized $J\times 1$ vector of parameters of the model $\mathcall{M}$, where $\mathbf z_j$ is a user-defined vector of $K\times 1$ (finite) points, whereas $\mathbf p_j$ is a $K\times 1$ vector unknown probabilities obeying to $\mathbf p_j^T\mathbf 1_K = 1$. Note that the arrays $\mathbf z_1,\ldots,\mathbf z_J$ must be chosen to cover the natural range of the model parameters. Thus, for instance, in the case of estimating the population mean $\mu \in \mathbb{R}$ for a normal model $\text{N}(\mu,\sigma^2)$ with $\sigma^2$ known, $\mathbf z_\mu = (-d,\ldots,0,\ldots,d)^T$ with $d$ as large as possible. In practice, as observations $\mathbf y$ are available, the support vector can be defined using sample information, i.e., $\mathbf z_\mu = \left(\min(\mathbf y),\ldots,\max(\mathbf y)\right)^T$. Similarly, in the case of estimating the parameter $\pi \in [0,1]$ of the Binomial model $\text{Bin}(\pi,n)$, the support vector is $\mathbf z_\pi = (0,\ldots,1)^T$. The choice of the number of points $K$ of $\mathbf z$ can be made via sensitivity analysis although it has been shown that $K\in \{5,7,11\}$ is usually enough for many regular problems (e.g., see \cite{papalia2008composite,golan1996maximum}). {Readers may refer to \cite{golan1996maximum} and \cite{ciavolino2014generalized} for further details.}  

Under the reparameterization in Equation \eqref{eq1}, $\mathcall U(\boldsymbol{\theta}) = \mathbf 0_J$ is solved via the following NLP problem:
\begin{equation}\label{eq2}
\begin{aligned}
&\underset{(\mathbf p_1,\ldots,\mathbf p_J)}{\text{maximize}} & \mathcall H(\mathbf{p}_1,\ldots,\mathbf{p}_J) \\
&\text{subject to:} & \mathcall U(\boldsymbol{\tilde\theta}) = \mathbf 0_J\\
&& \mathbf{p}_1^T\mathbf 1_K = 1\\
&& \vdots\\
&& \mathbf{p}_J^T\mathbf 1_K = 1,\\
\end{aligned}
\end{equation}
where $\mathcall H(\mathbf{p}) = -\sum_{j=1}^{J}\mathbf p_j^T\log \mathbf p_j$ is the Shannon's entropy function, whereas the score equation $\mathcall U(\boldsymbol{\tilde\theta})$ has been rewritten using the reparameterized parameters $\boldsymbol{\tilde\theta}$. The problem needs to recover $K\times J$ quantities which are defined in a (convex) hyper-simplex region with $J$ (non-) linear equality constraints $\mathcall U({\tilde{\theta}_1}),\ldots,\mathcall U({\tilde{\theta}_J})$ (\textit{consistency constraints}) and linear equality constraints $\mathbf{p}_1^T\mathbf 1_K,\ldots,\mathbf{p}_J^T\mathbf 1_K$ (\textit{normalization constraints}). The latter ensure that the recovered quantities $\mathbf{\hat p}_1,\ldots,\mathbf{\hat p}_J$ are still probabilities. Note that closed-form solutions for the ME-score problem do not exist and solutions need to be attained numerically. 

In the following examples, we will show how the ME-score problem can be formulated in the most simple cases of estimating a mean from Normal, Poisson, and Gamma models (Examples 1-3) as well as in more complex cases of estimating parameters for logistic regression (Example 4).

\subsection{Example 1: The Normal case}\label{ex1}

Consider the case of estimating the location parameter $\mu \in \mathbb R$ of a Normal density function with $\sigma^2$ known. In particular, let 
$$\mathbf y = (2.61,4.18,3.40,3.73,3.63,2.41,3.76,3.93,4.66,1.59,4.51,2.77)^T$$ 
be a sample of $n=12$ drawn from a population with Normal density $\text{N}(\mu,\sigma_0^2)$ with $\sigma_0^2 = 1$ known. Our objective is to estimate $\mu$ using the information of $\mathbf y$. Let $$l(\mu) = (\sigma_0^2)^{-1}|| \mathbf y - \mu\mathbf 1_n ||^2 $$ be the log-likelihood of the model where constant terms have been dropped and $$\mathcall U(\mu) = (\sigma_0^2)^{-1}\left(\mathbf y^T\mathbf 1_n - n\mu\right)$$ be the corresponding score w.r.t. $\mu$. To define the associated ME-score problem to solve $\mathcall U(\mu) = 0$, first let ${\mu}_{\text{ME}} = \mathbf z^T \mathbf p$ with $\mathbf z$ and $\mathbf p$ being $K\times 1$ vector of supports and unknown probabilities. In this example, 
$$\mathbf z = \left(1.59,2.10,2.61,3.13,3.64,4.15,4.66\right)^T$$ 
with $K=7$, $z_1 = \min(\mathbf y)$, and $z_K = \max(\mathbf y)$. Given the optimization problem in \eqref{eq2}, in this case $\mathbf p$ can be recovered via the Lagrangean method, as follows. Let

\begin{equation}\label{eq3}
\mathcall L(\lambda_0,\lambda_1,\mathbf p) = -\mathbf p^T\log \mathbf p - \lambda_0\left(1 - \mathbf{p}^T\mathbf 1_K\right) - \lambda_1\left((\sigma_0^2)^{-1}(\mathbf y^T\mathbf 1_n - n(\mathbf z^T\mathbf p))\right)
\end{equation}
be the Lagrangean function, with $\lambda_0$ and $\lambda_1$ being the usual Lagrangean multipliers. The Lagrangean system of the problem is

\begin{align}
& \frac{\partial \mathcall L(\lambda_0,\lambda_1,\mathbf p)}{\partial \mathbf p} = -\log(\mathbf p) -1 -\lambda_0 - \lambda_1 n \mathbf z = \mathbf 0_K\label{eq4a}\\
& \frac{\partial \mathcall L(\lambda_0,\lambda_1,\mathbf p)}{\partial \lambda_0} = 1 - \mathbf{p}^T\mathbf 1_K = 0\label{eq4b}\\
& \frac{\partial \mathcall L(\lambda_0,\lambda_1,\mathbf p)}{\partial \lambda_1} = (\sigma_0^2)^{-1}(\mathbf y^T\mathbf 1_n - n(\mathbf z^T\mathbf p) = 0.\label{eq4c}
\end{align}

\noindent Solving $\mathbf p$ in Equation \eqref{eq4a}, by using Equation \eqref{eq4c}, we get the general solutions for the ME-score problem:

\begin{equation}\label{eq5}
\mathbf{\hat p} = \frac{\exp\left(-\mathbf z \hat\lambda_1 n (\sigma_0^2)^{-1}\right)}{\exp\left(-\mathbf z \hat\lambda_1 n (\sigma_0^2)^{-1}\right)^T\mathbf 1_K},
\end{equation}

\noindent where the quantity in the denominator is the normalization constant. Note that solutions in Equation \eqref{eq5} depend on the Lagrangean multiplier $\hat\lambda_1$, which needs to be determined numerically \cite{golan1998maximum}. In this particular example, we estimate the unknown Lagrangean multiplier using a grid-search approach, yielding to $\hat\lambda_1 = -0.024$. The final solutions are
$$ \mathbf{\hat p} = \left(0.087,0.101,0.117,0.136,0.159,0.185,0.215\right)^T$$
with $\hat\mu_{\text{ME}} = \mathbf z^T\mathbf{\hat p} = 3.432$, which corresponds to the maximum likelihood estimate of $\hat\mu_{\text{ML}} = \frac{1}{n}\mathbf y^T\mathbf 1_n = 3.432$, as expected.

\subsection{Example 2: The Poisson case}

Consider the simple case of estimating $\lambda \in \mathbb R^+$ of a Poisson density function. Let 
$$\mathbf y = (5,7,7,4,4,8,15,7,7,4,7,3,8,5,4,7)^T$$
be a sample of $n=16$ drawn from a Poisson density $\text{Pois}(\lambda)$ and $\mathcall U(\lambda) = -n + (\mathbf y^T\mathbf 1_n)/\lambda$ be the score of the model. The reparameterized Poisson parameter is $\lambda_{\text{ME}} = \mathbf z^T\mathbf p$, with support being defined as follows:
$$
\mathbf z = \left(0.00, 3.75, 7.50, 11.25, 15.00\right)^T,
$$
where $K=5$ and $z_K = \max(\mathbf y)$. Note that, since the Poisson parameter $\lambda$ is bounded below by zero, we can set $z_1 = 0$. Unlike the previous case, we cannot determine $\mathbf{\hat p}$ analytically. For this reason, we need to solve the ME-score problem: 
\begin{equation}
\begin{aligned}
&\underset{\mathbf p}{\text{maximize}} & -\mathbf p^T\log (\mathbf p) \label{eq6}\\
& \text{subject to:} & \mathbf p^T\mathbf 1_{K} \\
&& \mathbf -n + (\mathbf y^T\mathbf 1_n)/(\mathbf z^T\mathbf p) \\
\end{aligned}
\end{equation}
via the augmented Lagrangean adaptive barrier algorithm as implemented in the function \texttt{constrOptim.nl} of the \texttt{R} package \texttt{alabama} \cite{alabama2015}. The algorithm converged successfully in few	 iterations. The recovered probabilities are as follows:
$$\mathbf{\hat p} = \left(0.184,0.256,0.283,0.247,0.034\right)^T$$
with $\hat{\lambda}_{\text{ME}} = 6.375$, which is equal to the maximum likelihood solution $\hat{\lambda}_{\text{ML}} = \frac{1}{n}\mathbf y^T\mathbf 1_n = 6.375$, as expected.

\subsection{Example 3: The Gamma case}

Consider the following random sample $$\mathbf y = (0.09,0.35,0.98,0.20,0.44,0.13,0.25,0.48,0.09,0.45,0.03,0.06,0.18,0.26,0.79,0.36,0.26)^T$$ drawn from a Gamma density $\text{Ga}(\alpha,\rho)$ with $\alpha \in \mathbb R^+$ being the scale parameter and $\rho \in \mathbb R^+$ the rate parameter. The log-likelihood of the model is as follows: $$l(\alpha,\rho) = -((\alpha-1)\log(\mathbf y)^T\mathbf{1}_n - (\mathbf y^T\mathbf 1_n\rho) + n\alpha\log(\rho) - n\log\left(\Gamma(\alpha))\right)$$ where $\Gamma(.)$ is the well-known gamma function. The corresponding score function equals to 
\begin{align*}
& \mathcal U(\alpha) = -\mathbf y^T\mathbf 1_n + {n\alpha}{\rho^{-1}}\\
& \mathcal U(\rho) = \log(\mathbf y)^T\mathbf{1}_n + n\log(\rho) - n\psi(\alpha)	
\end{align*}
with $\psi(\alpha) = \frac{\partial}{\partial\alpha} \log(\Gamma(\alpha))$ being the digamma function, i.e. the derivative of the logarithm of the Gamma function evaluated in $\alpha$. The re-parameterized Gamma parameters are defined as usual $\tilde{\alpha}_{\text{ME}} = \mathbf z_\alpha^T \mathbf p_\alpha$ and $\tilde{\rho}_{\text{ME}} = \mathbf z_\rho^T \mathbf p_\rho$ whereas the supports can be determined as $\mathbf z_\alpha = \left(0,\ldots,\overline{\alpha}+\delta\right)$ and $\mathbf z_\rho = \left(0,\ldots,\overline{\rho}+\delta\right)$, with $\delta$ being a positive constant. Note that the upper limits of the support can be chosen according to the following approximations: $\overline{\alpha} = 1\big/2M$ and $\overline{\rho} = \overline{\alpha}\big/ \overline y$, with $M = \log(\overline{y}) - {\sum_i \log(y_i)}\big/{n}$ and $\overline{y} = \sum_i y_i \big/ n$ \cite{choi1969maximum}. In the current example, the supports for the parameters are:
$$
\mathbf z_\alpha = \left(0.00, 1.12, 2.24, 3.35, 4.47\right)^T \quad\text{ and }\quad \mathbf z_\rho = \left(0.00, 1.91, 3.82, 5.73, 7.64\right)^T
$$		
where $K=5$, $\overline{\alpha} = 1.47$, $\overline{\rho} = 4.64$, and $\delta = 3$. The ME-score problem for the Gamma case is
\begin{equation}
\begin{aligned}
&\underset{\mathbf p}{\text{maximize}} & -\mathbf p_\alpha^T\log (\mathbf p_\alpha) - \mathbf p_\rho^T\log (\mathbf p_\rho)  \label{eq6b}\\
& \text{subject to:} & \mathbf p_\alpha^T\mathbf 1_{K} \\
&& \mathbf p_\rho^T\mathbf 1_{K}\\
&& -\mathbf y^T\mathbf 1_n + ({n \mathbf z_\alpha^T \mathbf p_\alpha})({\mathbf z_\rho^T \mathbf p_\rho}){^{-1}} \\
&& \log(\mathbf y)^T\mathbf{1}_n + n\log(\mathbf z_\rho^T \mathbf p_\rho) - n\psi(\mathbf z_\alpha^T \mathbf p_\alpha)
\end{aligned}
\end{equation}
which is solved via Augmented Lagrangean adaptive barrier algorithm. The algorithm required few iterations to converge and the recovered probabilities are as follows:
$$
\mathbf{\hat p}_\alpha = (0.290, 0.261, 0.222, 0.164, 0.063)^T \quad\text{ and }\quad \mathbf{\hat p}_\rho = (0.058, 0.138, 0.208, 0.270, 0.327)^T
$$
The estimated parameters under the ME-score formulation are $\hat{\alpha}_\text{ME} = 1.621$ and $\hat{\rho}_\text{ME} = 5.103$ which equal to the maximum likelihood solutions $\hat{\alpha}_\text{ML} = 1.621$ and $\hat{\rho}_\text{ML} = 5.103$.

\subsection{Example 4: Logistic regression}

In what follows, we show the ME-score formulation for logistic regression. We will consider both the cases of simple situations involving no separation---where maximum likelihood estimates can be easily computed---and those unfortunate situations in which separation occur. Note that in the latter case, maximum likelihood estimates are no longer available without resorting to the use of a bias reduction iterative procedure \cite{firth1993bias}. Formally, the logistic regression model with $p$ continuous predictors is as follows:
\begin{align}
& \pi_i = \left(1 + \exp(-\mathbf X_i\boldsymbol\beta)\right)^{-1}\label{eq7}\\
& y_i \sim \text{Bin}\left(\pi_i\right),\nonumber
\end{align}

\noindent where $\mathbf X$ is an $n\times p$ matrix containing predictors, $\boldsymbol\beta$ is a $p\times 1$ vector of model parameters, and $\mathbf y$ is an $n\times 1$ vector of observed responses. Here, the standard maximum likelihood solutions $\hat{\boldsymbol\beta}$ are usually attained numerically, e.g., using Newton--Raphson like algorithms \cite{albert1984existence}. \\

\noindent\textit{No separation case}. As an illustration of the ME-score problem in the optimal situation where no separation occurs, we consider the traditional Finney's data on vasoconstriction in the skin of the digits (see Table \ref{tab1}) \cite{pregibon1981logistic}.

\begin{table}[!h]
	\caption{Finney's data on vasoconstriction in the skin of the digits. The response $Y$ indicates the occurrence ($Y=1$) or non-occurrence ($Y=0$) of the vasoconstriction.}
	\label{tab1}
	\centering
	\begin{tabular}{ccc}
		\toprule
		Volume&Rate&Y\\
		\midrule
		3.70&0.825&1\\
		3.50&1.090&1\\
		1.25&2.500&1\\
		0.75&1.500&1\\
		0.80&3.200&1\\
		0.70&3.500&1\\
		0.60&0.750&0\\
		1.10&1.700&0\\
		0.90&0.750&0\\
		0.90&0.450&0\\
		0.80&0.570&0\\
		0.55&2.750&0\\
		0.60&3.000&0\\
		1.40&2.330&1\\
		0.75&3.750&1\\
		2.30&1.640&1\\
		3.20&1.600&1\\
		0.85&1.415&1\\
		1.70&1.060&0\\
		\bottomrule
	\end{tabular}
\end{table}

\noindent In the Finney's case, the goal is to predict the vasoconstriction responses as a function of Volume and Rate, according to the following linear term \cite{pregibon1981logistic}:
\begin{equation}\label{eq8a}
\text{logit}(\pi_i) = \beta_0 + \beta_1\log{\left(\text{Volume}_i\right)} + \beta_2\log{\left(\text{Rate}_i\right)}
\end{equation}

\noindent with $\text{logit}:[0,1]\to \mathbb{R}$ being the inverse of the logistic function. In the maximum entropy framework, the model parameters can be reformulated as follows:
\begin{align}\label{eq8b}
& \boldsymbol\beta_{\text{ME}} = \left(\mathbf z^T \otimes \mathbf I_{p+1}\right)\text{vec}(\mathbf P^T),
\end{align}

\noindent where $\mathbf z$ is a $K\times 1$ vector of support points, $\mathbf I_{p+1}$ is an identity matrix of order $p+1$ (including the intercept term), $\mathbf P$ is a $(p+1)\times K$ matrix of probabilities associated to the $p$ parameters plus the intercept, $\otimes$ is the Kronecker product, whereas $\text{vec(~)}$ is a linear operator that transforms a matrix into a column vector. Note that in this example $p=2$ and $K=7$, whereas the support $\mathbf z = (-10,\ldots,0,\ldots,10)^T$ is defined to be the same for both predictors and the intercept (the bounds of the support have been chosen to reflect the maximal variation allowed by the logistic function). Finally, the ME-score problem for the Finney's logistic regression is:
\begin{equation}
\begin{aligned}
&\underset{\text{vec}(\mathbf P)}{\text{maximize}} & -\text{vec}(\mathbf P)^T\log (\text{vec}(\mathbf P)) \label{eq8c}\\
& \text{subject to:} & \text{vec}(\mathbf P)^T\mathbf 1_{p(K+1)} \\
&& \mathbf X^T(\mathbf y-\boldsymbol\pi), \\
\end{aligned}
\end{equation}
where $\mathbf X$ is the $n\times (p+1)$ matrix containing the variables Rate, Volume, and a column of all ones for the intercept term, and $\boldsymbol\pi = \left(1 + \exp(-\mathbf X \boldsymbol\beta_{\text{ME}})\right)^{-1}$, with $\boldsymbol\beta_{\text{ME}}$ being defined as in Equation \eqref{eq8b}. Solutions for $\mathbf{\hat P}$ were obtained via the augmented Lagrangean adaptive barrier algorithm, which yielded the following estimates: 
$$
\mathbf{\hat P} = 
\begin{bmatrix}
0.000&0.004&0.062&0.159&0.220&0.263&0.293\\
0.000&0.001&0.099&0.178&0.224&0.247&0.251\\
0.205&0.201&0.190&0.170&0.137&0.085&0.013
\end{bmatrix},
$$
where the third line of $\mathbf{\hat P}$ refers to the intercept term. The final estimated coefficients are
\begin{align*}
& \hat{\beta}_{0_{\text{ME}}} = -2.875\\
& \hat{\beta}_{1_{\text{ME}}} = 5.179\\
& \hat{\beta}_{2_{\text{ME}}} = 4.562,
\end{align*}
which are the same as those obtained in the original paper of \cite{pregibon1981logistic}. \\

\noindent\textit{Separation case}. As a typical example of data under separation, we consider the classical Fisher iris dataset \cite{lesaffre1989partial}. As generally known, the dataset contains fifty measurements of length and width (in centimeters) of sepal and petal variables for three species of iris, namely setosa, versicolor, and virginica \cite{fisher1936use}. 
For the sake of simplicity, we keep a subset of the whole dataset containing two species of iris (i.e., setosa and virginica) with sepal length and width variables only. Inspired by the work of \cite{lesaffre1989partial}, we study a model where the response variable is a binary classification of iris, with $Y=0$ indicating the class virginica and $Y=1$ the class setosa, whereas petal length and width are predictors of $Y$. The logistic regression for the iris data assumes the following linear term:
\begin{equation}\label{eq8d}
\text{logit}(\pi_i) = \beta_0 + \beta_1\text{length}_i + \beta_1\text{width}_i,
\end{equation}
where model parameters can be reformulated as in Equation \eqref{eq8b}, with $K=7$, $p=2$, and $\mathbf z$ being centered around zero with bounds $z_1 = -25$ and $z_K = 25$. The ME-score problem for the iris dataset is the same as in \eqref{eq8c} and it is solved using the augmented Lagrangean adaptive barrier algorithm. The recovered $\mathbf{\hat P}$ is
$$
\mathbf{\hat P} = 
\begin{bmatrix}
0.228& 0.226& 0.215& 0.190& 0.137& 0.001& 0.001\\
0.000& 0.039& 0.040& 0.158& 0.218& 0.257& 0.285\\
0.000& 0.000& 0.000& 0.037& 0.210& 0.329& 0.426
\end{bmatrix},
$$
where the intercept term is reported in the third line of the matrix. The estimates for the model coefficients are reported in Table \ref{tab2} (ME, first column). For the sake of comparison, Table \ref{tab2} also reports the estimates obtained by solving the score of the model via Bias-corrected Newton--Raphson (NRF, second column) and Newton--Raphson (NR, third column). The NRF algorithm uses the Firth's correction for the score function \cite{firth1993bias} as implemented in the \texttt{R} package \texttt{logistf} \cite{logistf2018}. As expected, the NR algorithm fails to converge reporting divergent estimates. By contrast, the NRF procedure converges to non-divergent solutions. Interestingly, the maximum entropy solutions are more close to NRF estimates although they differ in magnitude. 

\begin{table}[!h]
	\caption{Estimates for the iris logistic regression: ME (maximum entropy), NRF (Biased-corrected Newton--Raphson), NR (Newton--Raphson). Note that NRF algorithm implements the Firth's bias correction \cite{firth1993bias}.}
	\label{tab2}
	\centering
	\begin{tabular}{cccc}
		\toprule
		&ME&NRF&NR\\
		\midrule
		$\beta_0$ & 17.892 &12.539  &445.917\\
		$\beta_1$ &-10.091 &-6.151 &-166.637\\
		$\beta_2$ &12.229  &6.890  &140.570\\
		\bottomrule
	\end{tabular}
\end{table}

\section{Simulation study}\label{simulation_study}

Having examined the ME-score problem with numerical examples for both simple and more complex cases, in this section, we will numerically investigate the behavior of the maximum entropy solutions for the most critical case of logistic regression under separation. \\

\noindent\textit{Design}. Two factors were systematically varied in a complete two-factorial design:
\begin{itemize}
	\item[(i)] the sample size $n$ at three levels: 15, 20, 200; 
	\item[(ii)] the number of predictors $p$ (excluding the intercept) at three levels: 1, 5, 10.
\end{itemize} 
The levels of $n$ and $p$ were chosen to represent the most common cases of simple, medium, and complex models, as those usually encountered in many social research studies. \\

\noindent\textit{Procedure}. Consider the logistic regression model as represented in Equation \eqref{eq7} and let $n_k$ and $p_k$ be distinct elements of sets $n$ and $p$. The following procedure was repeated $Q=10000$ times for each of the $n\times p = 9$ combinations of the simulation design:
\begin{enumerate}
	\item Generate the matrix of predictors $\mathbf X_{n_k \times (1+p_k)} = \left[\mathbf 1_{n_k} | \tilde{\mathbf X}_{n_k\times p_k}\right]$, where $\tilde{\mathbf X}_{n_k\times p_k}$ is drawn from the multivariate standard Normal distribution $\text{N}(\mathbf{0}_{p_k},\mathbf{I}_{p_k})$, whereas the column vector of all ones $\mathbf 1$ stands for the intercept term;
	\item Generate the vector of predictors $\boldsymbol\beta_{1+p_k}$ from the multivariate centered Normal distribution $\text{N}(\mathbf{0}_{1+p_k},\sigma\mathbf{I}_{1+p_k})$, where $\sigma = 2.5$ was chosen to cover the natural range of variability allowed by the logistic equation;
	\item Compute the vector $\boldsymbol\pi_{n_k}$ via Equation \eqref{eq7} using $\mathbf X_{n_k \times (1+p_k)}$ and $\boldsymbol\beta_{p_k}$;
	\item For $q=1,\ldots,Q$, generate the vectors of response variables $\mathbf y_{n_k}^{(q)}$ from the binomial distribution $\text{Bin}(\boldsymbol\pi_{n_k})$, with $\boldsymbol\pi_{n_k}$ being fixed;
	\item For $q=1,\ldots,Q$, estimate the vectors of parameters $\hat{\boldsymbol\beta}_{1+p_k}^{(q)}$ by means of Newton--Raphson (NR), Bias-corrected Newton--Raphson (NRF), and ME-score (ME) algorithms.
\end{enumerate}
The entire procedure involves a total of $10000 \times 3 \times 3 = 90000$ new datasets as well as an equivalent number of model parameters. For the NR and NRF algorithms, we used the \texttt{glm} and \texttt{logistf} routines of the \texttt{R} packages \texttt{stats} \cite{Rlang} and \texttt{logistf} \cite{logistf2018}. By contrast, the ME-score problem was solved via the augmented Lagrangean adaptive barrier algorithm implemented in \texttt{constrOptim.nl} routine of the \texttt{R} package \texttt{alabama} \cite{alabama2015}. Convergences of the algorithms were checked using the built-in criteria of \texttt{glm}, \texttt{logistf}, and \texttt{constrOptim.nl}. For each of the generated data $\{\mathbf y, \mathbf X\}_{q=1,\ldots,Q}$, the occurrence of separation was checked using a linear programming-based routine to find infinite estimates in the maximum likelihood solution \cite{Konis2007,Konis2013}. The whole simulation procedure was performed on a (remote) HPC machine based on 16 cpu Intel Xeon CPU E5-2630L v3 1.80GHz, 16x4GB Ram.\\

\noindent\textit{Measures}. The simulation results were evaluated considering the averaged bias of the parameters $\hat{B} = \frac{1}{Q}(\boldsymbol\beta^{(k)}-\hat{\boldsymbol\beta}^{(k)})^T\mathbf 1$, its squared version $\hat{B}^2$ (the square here is element-wise), and the averaged variance of the estimates $\hat{V} = \frac{1}{Q}\text{Var}(\hat{\boldsymbol\beta}^{(k)})$. They were then combined together to form the mean square error (MSE) of the estimates $\text{MSE} = \hat{V} + \hat{B}^2 $. {The relative bias $RB = (\hat{\beta}_j^{(k)} - \beta^0_j)\big/|\beta^0_j|$ was also computed for each predictor $j=1,\ldots,J$, ($\beta^0$ indicates the population parameter).} The measures were computed for each of the three algorithms and for all the combinations of the simulation design.\\

\begin{figure}[!h]
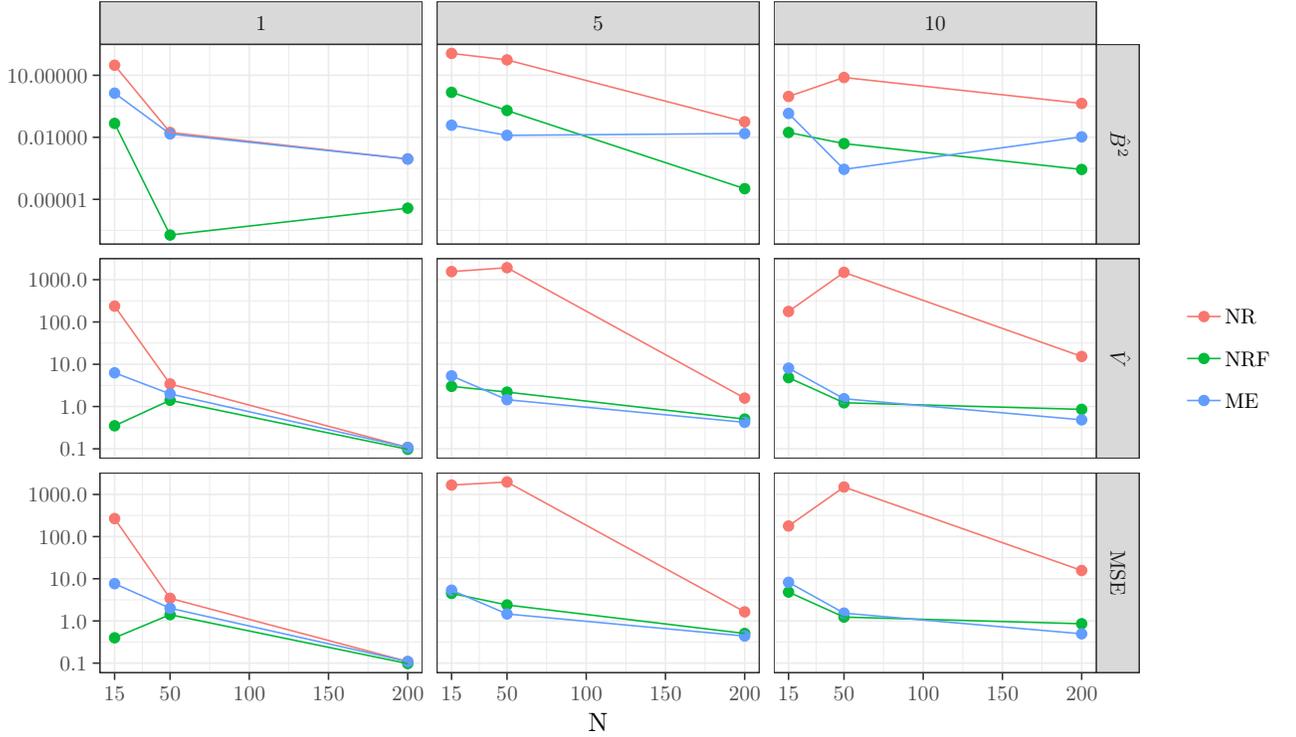

	\include{biasmedi}
	\caption{Simulation study: averaged bias, squared averaged bias, and MSE for NR, NRF, ME algorithms. Note that the number of predictors $p$ is represented column-wise (outside) whereas the sample sizes $n$ is reported in the x-axis (inside). The measures are plotted on logarithmic scale.}
	\label{fig1}
\end{figure}

\begin{figure}[!h]
	\hspace*{-1cm}\scalebox{.97}{\input{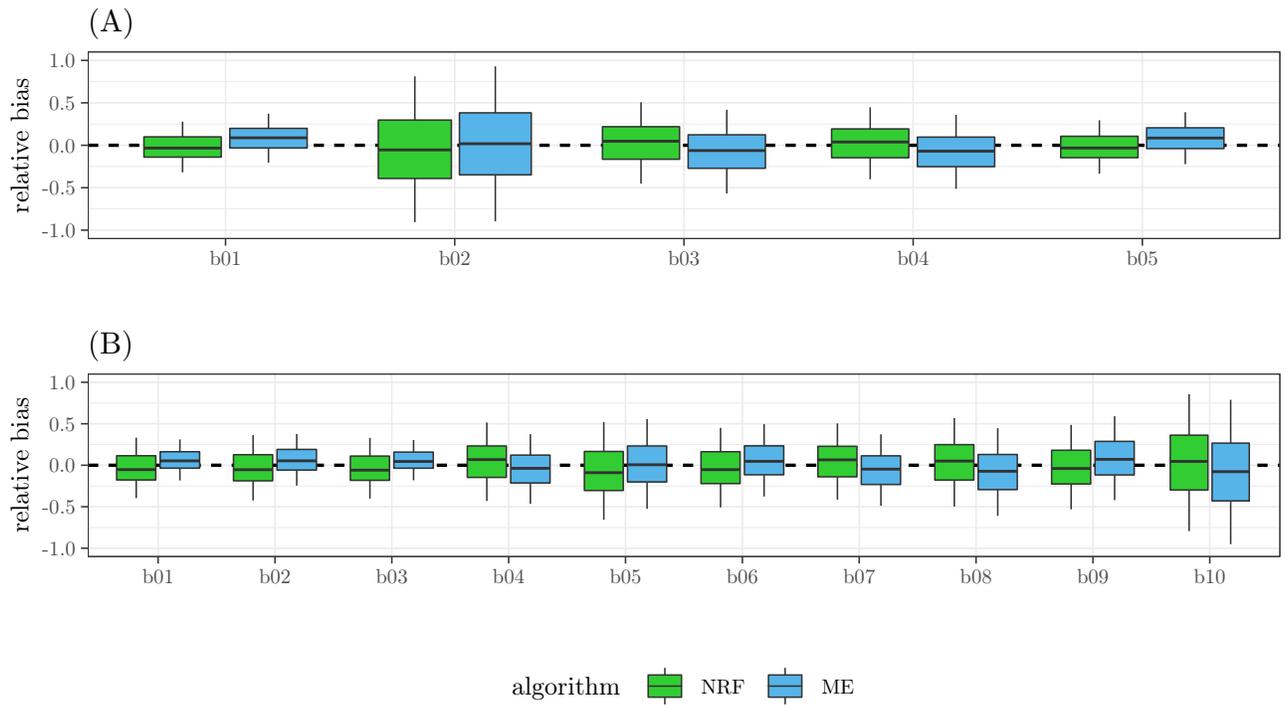}}
	\caption{Simulation study: relative bias for NRF and ME algorithms in the conditions $N=200 \land P=5$ (A) and $N=200 \land P=10$ (B). Note that plots are paired vertically by predictor. Rate of over-estimation (under-estimation): (A) ME = 0.54 (0.46), NRF = 0.49 (0.51); (B) ME = 0.53 (0.47), NRF = 0.47 (0.53). }
	\label{fig2}
\end{figure}

\noindent\textit{Results}. Table \ref{tab3} reports the proportions of separation present in the data for each level of the simulation design along with the proportions of non-convergence for the three algorithms. As expected, NR failed to converge when severe separation occurred, for instance, in the case of small samples and large number of predictors. By contrast, for NRF and ME algorithms, the convergence criteria were always met. The results of the simulation study with regards to bias, variance, and mean square error (MSE) are reported in Table \ref{tab4} and Figure \ref{fig1}. In general, MSE for the three algorithms decreased almost linearly with increasing sample sizes and number of predictors. As expected, the NR algorithm showed higher MSE than NRF and ME, except in the simplest case of $n=200$ and $p=1$. Unlike for the NR algorithm, with increasing model complexity ($p>1$), ME showed a similar performances of NRF both for medium ($n=50$) and large ($n=200$) sample sizes. Interestingly, for the most complex scenario, involving a large sample ($n=200$) and higher model complexity ($p=10$), the ME algorithm outperformed NRF in terms of MSE. {To further investigate the relationship between NRF and ME, we focused on the latter conditions and analyzed the behavior of ME and NRF in terms of relative bias (RB, see Figure \ref{fig2}). Both the ME and NRF algorithms showed RB distributions centered about 0. Except for the condition $N=200 \land P=10$, where ME showed smaller variance than NRF, both the algorithms showed similar variance in the estimates of the parameters. Finally, we also computed the ratio of over- and under-estimation $r$ as the ratio between the number of positive RB and negative RB, getting the following results: $r_{\text{ME}}=1.18$ (over-estimation: 54\%), $r_{\text{NRF}}=0.96$ (over-estimation: 49\%) for the case $N=200 \land P=5$ and $r_{\text{ME}}=1.12$ (over-estimation: 53\%), $r_{\text{NRF}}=0.91$ (over-estimation: 47\%) for the case $N=200 \land P=10$. \\
	Overall, the results suggest the following points:
	\begin{itemize}
		\item In the simplest cases with no separation (i.e., $N=50 \land P=1$, $N=200 \land P=1$, $N=200 \land P=5$), the ME solutions to the maximum likelihood equations were the same as those provided by standard Newton--Raphson (NR) and the bias-corrected version (NRF). In all these cases, the bias of the estimates approximated zero (see Table \ref{tab4});
		\item In the cases of separation, ME showed comparable performances to NRF, which is known to provide the most efficient estimates in the case of logistic model under separation: Bias and MSE decreased as a function of sample size and predictors, with MSE being lower for ME than NRF in the case of $N=200 \land P=5$ and $N=200 \land P=10$; 
		\item In the most complex scenario with a large sample and higher model complexity ($N=200 \land P=5$, $N=200 \land P=10$), ME and NRF algorithms showed similar relative bias, with ME estimates being less variable than NRF in $N=200 \land P=10$ condition. The ME algorithm tended to over-estimate the population parameters, by contrast NRF tended to under-estimate the true model parameters. 
	\end{itemize}
}

\begin{table}[!h]
	\caption{Simulation study: proportions of separation occurred in the data and non-convergence (nc) rates for NR, NRF, ME algorithms.}
	\label{tab3}
	\centering
	\begin{tabular}{cccccc}
		\toprule
		n & p & separation & $\text{nc}_\text{\tiny NR}$ & $\text{nc}_\text{\tiny NRF}$ & $\text{nc}_\text{\tiny ME}$ \\ 
		\midrule
		15 & 1 & 0.333 & 0.085 & 0.000 & 0.000 \\ 
		50 & 1 & 0.002 & 0.002 & 0.000 & 0.000 \\ 
		200 & 1 & 0.000 & 0.000 & 0.000 & 0.000 \\ 
		15 & 5 & 0.976 & 0.237 & 0.000 & 0.000 \\ 
		50 & 5 & 0.771 & 0.771 & 0.000 & 0.000 \\ 
		200 & 5 & 0.000 & 0.000 & 0.000 & 0.000 \\ 
		15 & 10 & 1.000 & 0.002 & 0.000 & 0.000 \\ 
		50 & 10 & 0.949 & 0.950 & 0.000 & 0.000 \\ 
		200 & 10 & 0.013 & 0.013 & 0.000 & 0.000 \\ 
		\bottomrule
	\end{tabular}
\end{table}

\begin{table}[!h]
	\caption{Simulation study: averaged bias, squared averaged bias, and MSE for NR, NRF, ME algorithms.}
	\label{tab4}
	\centering
	\begin{tabular}{ll|rrrr|rrrr|rrrr}\toprule
		&  & \multicolumn{4}{|c|}{NR} & \multicolumn{4}{c|}{NRF} & \multicolumn{4}{c}{ME} \\ \midrule
		n & p &  $\hat B$ & $\hat V$ & $\hat{B}^2$ & $MSE$ &  $\hat B$ & $\hat V$ & $B^2$ & $MSE$ &  $\hat B$ & $\hat V$ & $B^2$ & $MSE$  \\ \hline
		15 & 1 & -5.54 & 236.70 & 30.67 & 267.36 & 0.22 & 0.35 & 0.05 & 0.40 & -1.17 & 6.28 & 1.37 & 7.64 \\ 
		50 & 1 & -0.13 & 3.42 & 0.02 & 3.44 & -0.00 & 1.41 & 0.00 & 1.41 & -0.12 & 1.99 & 0.01 & 2.00 \\ 
		200 & 1 & 0.03 & 0.11 & 0.00 & 0.11 & 0.00 & 0.10 & 0.00 & 0.10 & 0.03 & 0.11 & 0.00 & 0.11 \\ 
		&&&&&&&&&&&&&\\
		15 & 5 & 10.68 & 1553.37 & 113.98 & 1667.33 & -1.22 & 3.00 & 1.50 & 4.49 & 0.20 & 5.32 & 0.04 & 5.36 \\ 
		50 & 5 & 7.46 & 1918.18 & 55.65 & 1973.78 & -0.44 & 2.20 & 0.20 & 2.39 & -0.11 & 1.45 & 0.01 & 1.46 \\ 
		200 & 5 & 0.24 & 1.58 & 0.06 & 1.64 & 0.01 & 0.50 & 0.00 & 0.50 & 0.12 & 0.42 & 0.02 & 0.44 \\ 
		&&&&&&&&&&&&&\\
		15 & 10 & -0.97 & 177.40 & 0.95 & 178.35 & -0.13 & 4.82 & 0.02 & 4.84 & -0.38 & 8.10 & 0.14 & 8.24 \\ 
		50 & 10 & 2.80 & 1490.39 & 7.83 & 1498.20 & -0.07 & 1.23 & 0.00 & 1.23 & -0.02 & 1.53 & 0.00 & 1.53 \\ 
		200 & 10 & 0.66 & 15.29 & 0.43 & 15.72 & 0.02 & 0.86 & 0.00 & 0.86 & 0.10 & 0.48 & 0.01 & 0.50 \\ 
		\bottomrule
	\end{tabular}
\end{table}


\section{Discussion and conclusion}\label{discussion}

We have described a new approach to solve the problem $\mathcal U(\boldsymbol\theta)=\mathbf 0$ in order to get $\boldsymbol{\hat{\theta}}$ in the context of maximum likelihood theory. Our proposal took the advantages of using the maximum entropy principle to set a non-linear programming problem where $\mathcal U(\boldsymbol\theta)$ was not solved directly, but it was used as informative constraint to maximize the Shannon's entropy. Thus, the parameter $\boldsymbol\theta$ was not searched over the parameter space $\boldsymbol\Theta\subset\mathbb R^J$, rather it was reparameterized as a convex combination of a known vector $\mathbf z$, which indicated the finite set of possible values for $\boldsymbol\theta$, and a vector of unknown probabilities $\mathbf p$, which instead needed to be estimated. In so doing, we converted the problem $\mathcal U(\boldsymbol\theta)=\mathbf 0$ from one of numerical mathematics to one of inference, where $\mathcal U(\boldsymbol\theta)$ was treated as one of the many pieces of (external) information we may have had. As a result, the maximum entropy solution did not require either the computation of the Hessian of second-order derivatives of $l(\boldsymbol\theta)$ (or the expectation of the Fisher information matrix) or the definition of initial values, as is required by Newton-like algorithms $\boldsymbol\theta^0$. In contrast, the maximum entropy solution revolved around the reduction of the initial uncertainty: as one adds pieces of external information (constraints), a departure from the initial uniform distribution $\mathbf p$ results, implying a reduction of the uncertainty about $\boldsymbol\theta$; a solution is found when no further reduction can be enforced given the set of constraints. We used a set of empirical cases and a simulation study to assess the maximum entropy solution to the score problem. In cases where the Newton--Raphson is no longer correct for $\boldsymbol\theta$ (e.g., logistic regression under separation), the ME-score formulation showed results (numerically) comparable with those obtained using the Bias-corrected Newton--Raphson, in the sense of having the same or even smaller mean square errors (MSE). Broadly speaking, these first findings suggest that the ME-score formulation can be considered as a valid alternative to solve $\mathcal U(\boldsymbol\theta)=\mathbf 0$, although further in-depth investigations need to be conducted to formally evaluate the statistical properties of the ME-score solution.

Nevertheless, we would like to say that the maximum entropy approach has been used to build a solver for maximum likelihood equations \cite{fang2012entropy,el2003maximum,sukumar2004construction}. In this sense, standard errors, confidence levels, and other likelihood based quantities can be computed using the usual asymptotic properties of maximum likelihood theory. However, attention should be directed at the definition of the support points $\mathbf z$ since they need to be sufficiently large to include the true (hypothesized) parameters we are looking for. Relatedly, our proposal differs from other methods, such as Generalized Maximum Entropy (GME) or Generalized Cross Entropy (GCE) \cite{golan1996maximum,ciavolino2016generalized}, in two important respects. First, the ME-score formulation does not provide a class of estimators for the parameters of statistical models. By contrast, GME and GCE are estimators belonging to the exponential family, which can be used in many cases as alternatives to maximum likelihood estimators \cite{golan1997maximum}. Second, the ME-score formulation does not provide an inferential framework for $\boldsymbol\theta$. While GME and GCE use information theory to provide the basis for inference and model evaluation (e.g., using Lagrangean multipliers and normalized entropy indices), the ME-score formulation focuses on the problem of finding roots for $\mathcal U(\boldsymbol\theta)=\mathbf 0$. {Finally, an open issue which deserves greater consideration in future investigations is the examination of how the ME-score solution can be considered in light of the well-known maximum entropy likelihood duality \cite{brown1986fundamentals}.}

Some advantages of the ME-score solution over Newton-like algorithms may include the following: (i) model parameters are searched in a smaller and simpler space because of the convex reparameterization required for $\boldsymbol\theta$; (ii) the function to be maximized does not require either the computation of second-order derivatives of $l(\boldsymbol\theta)$, or searching for good initial values $\boldsymbol\theta^0$; (iii) additional information on the parameters, such as dominance relations among the parameters, can be added to the ME-score formulation in terms of inequality constraints (e.g., $\boldsymbol\theta_{j} < \boldsymbol\theta_{t}$, $j\neq t$). Furthermore, the ME-score formulation may be extended to include a priori probability distributions on $\boldsymbol\theta$. While in the current proposal, the elements of $\mathbf z_j$ have the same probability to occur, the Kullback--Leibler entropy might be used to form a Kullback Leibler
-score problem, where $\mathbf z = (\mathbf z_1,\ldots,\mathbf z_J)^T$ are adequately weighted by known vectors of probability $\mathbf w = (\mathbf w_1,\ldots,\mathbf w_J)^T$. This would offer, for instance, another opportunity to deal with cases involving penalized likelihood estimations. 

In conclusion, we think that this work yielded initial findings in the solution of likelihood equations from a maximum entropy perspective. To our knowledge, this is the first time that maximum entropy is used to define a solver to the score function. We believe this contribution will be of interest to all researchers working at the intersection of information theory, data mining, and applied statistics.

\clearpage
\bibliographystyle{plain}
\bibliography{biblio}

\end{document}